\documentstyle[aps]{revtex}
\tightenlines
\begin{document}
\preprint{Evolutionary implications of a power-law
distribution of protein family sizes/Bader}
\title{Evolutionary implications of a power-law distribution of protein family sizes}

\author{Joel~S.~Bader\cite{jsb}}

\date{\today}
\maketitle

\noindent Corresponding Author

Joel S. Bader, CuraGen, 555 Long Wharf Drive, New Haven, CT,
06511.

Tel. (203)401-3330x236; Fax (203)401-3351; Email
jsbader@curagen.com

\clearpage

\begin{abstract}
Current-day genomes bear the mark of the evolutionary processes.
One of the strongest indications is the sequence homology among
families of proteins that perform similar biological functions in
different species.  The number of proteins in a family can grow
over time as genetic information is duplicated through evolution.
We explore how evolution directs the size distribution of these
families. Theoretical predictions for family sizes are obtained
from two models, one in which individual genes duplicate and a
second in which the entire genome duplicates.  Predictions from
these models are compared with the family size distributions for
several organisms whose complete genome sequence is known.  We
find that protein family size distributions in nature follow a
power-law distribution.  Comparing these results to the model
systems, we conclude that genome duplication is the dominant
mechanism leading to increased genetic material in the species
considered.
\end{abstract}

\clearpage

\narrowtext

\section{Introduction}

Current-day genomes are the result of generations of evolution.
One of the marks of evolution is the existence of protein
families. These families comprise groups of proteins that share
sequence similarity and perform similar biological functions
\cite{cog,gene_families}. The most likely explanation for the
similarity in sequence and function is that all the proteins in a
family evolved from a single common ancestor.

The size of a family, defined here as the number of proteins in a
family for a particular species, evolves over time through
processes that increase the physical size of an organism's genome.
Genomes in many major lineages are thought to have undergone
ancient doublings one or more times \cite{doubling}. It is thought
that genome doubling can provide an evolutionary advantage by
permitting redundant genes to evolve rapidly and perform different
biological roles, potentially allowing entire pathways to acquire
more specific function \cite{more_doubling}.

At finer scales, chromosomal regions or individual genes may be
may be duplicated or lost through evolution. Even without physical
loss, protein coding regions may suffer loss of function and cease
to be expressed, leading to the existence of pseudogenes
\cite{molecular_evolution}.

Previous studies have detected patterns supporting growth and loss
of genetic information.  Evolutionary processes consisting of
duplication and mutation can introduce long-range, power-law
correlations in the sequences of individual genes \cite{li};
reports of such correlations in intron-rich regions sparked
considerable interest \cite{stanley}.

In contrast to studies of individual gene sequences, we developed
a model to explain the evolution of the physical size of a genome
\cite{genome_size}.  In our model, a speciation rate allowed
genome size to increase or decrease, and an extinction rate
removed individual species. The ratio of the speciation and
extinction rates yielded scaling laws for the distribution of
genome sizes: exponential scaling when the amount of genetic
material lost or gained was constant, and power-law scaling
leading to a self-similar distribution when the change in genetic
material was proportional to the existing size.  Closed-form
approximations agreed with simulation results and explained
observations reported by others \cite{slonimski}.

Here we use related models to explore size of gene families.
Processes that add and remove genetic material are presented in
Sec. II. In the first model, we assume that duplication occurs on
the level of individual genes.  In the second model, we assume
that these events duplicate an entire genome. Closed-form
solutions are provided for the size distributions of gene
families. Next, in Sec. III, we present results from analysis of
gene families in sequenced genomes.  These results rely heavily on
the clusters of orthologous groups (COGs) database, which
identifies gene families that span eight individual unicellular
species including eubacteria, archaebacteria, cyanobacteria, and
eukaryots \cite{cog}.  We discuss which evolutionary model is most
consistent with our observations in Sec. IV.

\section{Theory}

For a single organism, let $P_n$ be the number of gene families
that contain $n$ genes.  The total number of families is $sum_{n}
P_n = P_{\rm tot}$. We describe two models for the increase or
decrease of the number of genes in the family.

\subsection{Model I: Gene Duplication}

In Model I, we assume that each gene in the family evolves independently.
Each gene duplicates with rate $k_+$ and each gene is lost
with rate $k_-$.  With each generation,
the change in the number of families of size $n$ is
\begin{equation}
\Delta P_n = (n-1) k_+ P_{n-1} + (n+1)k_- P_{n+1} - n( k_- + k_+) P_n.
\end{equation}
After sufficient time, the distribution reaches equilibrium
values. Detailed balance indicates that the number of families
increasing from size $n$ to $n+1$ should equal the number of
families decreasing from size $n+1$ to size $n$,
\begin{equation}
n k_+ P_n = (n+1)k_- P_{n+1}.
\end{equation}
The resulting expression for the populations is
\begin{equation}
P_n / P_{\rm tot} = (1/n) \alpha^n / [- ln( 1 - \alpha )],
\end{equation}
where we have defined $\alpha$ as $k_+/k_-$.  Alternatively,
normalizing by the families with a single member, we have
\begin{equation}
P_n/P_1 = (1/n) \alpha^{n - 1}.
\label{e:f1}
\end{equation}

In addition to describing dynamics when each gene is duplicated individually,
this model can also represent a system in which large genomic regions are
duplicated or lost, provided that only one member of the family is present
in the duplicated region.  If, for example, a single chromosome is duplicated,
this model could apply.

The populations $P_n/P_1$ predicted by Model I are shown as black lines
in Fig.~\ref{f:model} for
three choices of the parameter $\alpha$: 0.1 (thin black line),
0.3 (medium black line), and 0.9 (thick black line).  As the value of
$\alpha$ increases, the distribution of families shifts to larger sizes.
The shape of the distribution changes from a straight line on the log-log
plot at small $n$, characteristic of a power-law distribution, to a
curved line at larger $n$, characterstic of the faster decay of an
exponential distribution.

\subsection{Model II: Genome Duplication}

In Model II, we assume that genome duplication dominates the
evolutionary process.  Each genome can double in size with probability
$k_+$ or be reduced by half with probability $k_-$.  Writing the size of a
family after $j$ doublings as $n = 2^j$,
the evolution of $j$ at each generation is
\begin{equation}
\Delta P_j = k_+ P_{j - 1} + k_- P_{j+1} - (k_+ + k_-)P_j.
\end{equation}
Again relying on detailed balance, we find that $P_j \sim
\alpha^j$, with $\alpha = k_+/k_-$ as before.  For normalization,
we assume that $\sum_j P_j = P_{\rm tot}$, yielding
\begin{equation}
P_j = (1 - \alpha)\alpha^j.
\end{equation}

To change variables from $j$ to $n$, we make an approximation that
the discrete values of $j$ and $n$ may be replaced by a continuous
distribution. The distribution for $n$ is then $P_n =
P_{j(n)}dj(n)/dn$, where $j(n) = \log_2(n)$, giving
\begin{equation}
P_n / P_{\rm tot} = [(1-\alpha)/\ln 2] n^{(\ln \alpha / \ln 2) -
1}.
\end{equation}
Because we used a continuous distribution to derive this result,
the normalization is not exact.  The power-law form of the distribution,
however, is accurate, and simple summation may be used to define the
normalization constant.

Alternatively, the distribution may be defined relative to the number of
families of size 1, or
\begin{equation}
P_n / P_1 = n^{(\ln \alpha / \ln 2) - 1}. \label{e:f2}
\end{equation}
Results for $P_n/P_1$ are shown as grey lines in
Fig.~\ref{f:model} for three values of $\alpha$: 0.1 (thin grey
line), 0.3 (medium grey line), and 0.9 (thick grey line).  As
these are power-law distributions, they are straight on a log-log
plot.  The distribution favors larger family sizes as $\alpha$
increases.

\section{Results} \label{s:results}

To investigate the size distributions of gene families in nature,
we analyzed the contents of the COG database \cite{cog}.  This
database uses essentially unsupervised sequence-similarity
comparisons to group proteins into families of orthologs and
paralogs.  The current release includes 8328 proteins from eight
sequenced genomes (E. coli, H. influenzae, H. pylori, M.
genitalium, M. pneumoniaa, Synechocystis, M. jannaschii, and S.
cerevisiae) and assigns them to 864 individual families.  Only
proteins with orthologs in at least three species are included in
the database . Using this database, we computed the number of
families of size $n$, $P_n$, for each species, then normalized the
result by $P_1$ for the same species. The results of this analysis
are shown in Fig.~\ref{f:all}.

As seen in Fig.~\ref{f:all}, all the species show power-law
behavior for $P_n/P_1$ as a function of $n$ for families of size
10 or smaller.  The linear trend indicates that Model II,
duplication of the entire genome, is more likely than Model I, in
which individual genes are duplicated.

We explore the linear trend more quantitatively by performing a
least-squares fit of the data for each model. The quantity we
minimize is the RMS error for the log-transformed data,
\begin{equation}
{\rm RMS} = \sqrt{ (1/N)\sum_{n: P_n \geq 2} [
\log_{10}(P_n/P_1)_{\rm data} - \log_{10}(P_n/P_1)_{\rm fit} ]^2},
\end{equation}
with $(P_n/P_1)_{\rm fit}$ from Eq.~\ref{e:f1} or Eq.~\ref{e:f2}.
As noted in the summation, we considered only family sizes $n$
with $P_n = 2$ or more; the total number of family sizes used is
$N$. The results of the fit are detailed in Table~\ref{t:alpha},
along with the number of family sizes that contributed to the fit.
The model with the smaller RMS for the fit is also indicated.

As seen in Table~\ref{t:alpha}, Model II (complete genome
duplication) provides a consistently better fit to the data than
does Model I (individual gene duplication). In particular, when
all of the protein families for a given organism are considered,
each of the eight organisms shows a better fit with Model II than
with Model I.

In Table~\ref{t:alpha} the fit values for $\alpha$ are also shown
for the functional classes defined in the COG database:
information storage and processing, cellular processes,
metabolism, and poorly characterized \cite{cog}.  These individual
classes are also fit better by Model II than by Model I.  In E.
coli, H. influenzae, H. pylori, M. pnuemoniae, and Synechocystis,
at least three of the four classes are fit better by Model II; in
M. genitalium, there are not enough protein families for adequate
predictions of $\alpha$.  Only in S. cerevisiae does Model I
appear to provide a slightly better fit to the distribution of
family sizes for two classes, information storage and processing
and cellular processes.  One possible explanation for the better
performance of Model I for S. cerevisiae is that gene families
grow through the duplication of chromosomes, rather than the
duplication of individual genes or entire genomes.  The
distinction between the genome and individual chromosomes is not
applicable to the other organisms, which have a single chromosome.

A trend evident in Table~\ref{t:alpha} is that $\alpha$ for
cellular processes (molecular chaperones, outer membrane, cell
wall biogenesis, secretion, motility, inorganic ion transport and
metabolism) is typically larger than $\alpha$ for information
storage and processing (translation, ribosomal structure and
biogenesis, transcription, replication, repair, recombination) and
for metabolism (energy production and conversion, carbohydrate
metabolism and transport, amino acid metabolism and transport,
coenzyme metabolism, lipid metabolism).  Protein families for
cellular processes are therefore biased towards larger sizes,
while families for information storage and processing and
metabolism are biased toward smaller family sizes.  This would
imply that, in either model, a duplication of cellular process
proteins is more likely to be retained than duplications of other
functions.  This suggests that cells can tolerate changes to
cellular process pathways more readily than to other pathways.

The relative performance of Model I and Model II according to
protein family functional class is summarized in
Table~\ref{t:summary}. When all classes are considered, Model II
clearly provides a better explanation of the observed family
sizes.  When classes are considered separately, Model II provides
a better explanation for three classes (information storage and
processing, metabolism, and poorly characterized functions), while
Model I provides a better explanation only for cellular processes.

The fits provided by Model I and Model II are shown in
Fig.~\ref{f:fitorg} for E. coli and S. cerevisiae.  The observed
family size distributions are shown as points and the best fits as
lines, grey for Model I and black for Model II.  The top pair of
panels shows the results when all protein families are considered.
For families up to size 10, the distributions from both organisms
clearly follow the power-law prediction of Model II.

For the separate protein classes, the E. coli family sizes
continue to follow the power-law prediction of Model II.  As
mentioned previously for S. cerevisiae, however, the fit to Model
II is not good for the storage and processing and cellular
processes classes.  The size distribution decays much more rapidly
than Model II predicts.

\section{Discussion}

We have investigated the size distribution of protein families.
For a selection of single-celled organisms with sequenced genomes,
we find that the number of families with $n$ members follows a
power-law distribution as a function of $n$. This behavior
suggests that evolution increases protein diversity through
duplication of entire genomes, balanced occasionally by the loss
of large amounts of genetic information. It is less likely that
protein diversity is increased through the duplication of
individual genes, since this process would not lead to a power-law
distribution.

The power-law we find is that $P_n/P_1 ~ n^{-\alpha}$, where $P_n$
is the number of families of size $n$.  The exponent $\alpha$
varies from $0.2$ to $0.6$ depending on species.  In our theory,
this exponent measures the ratio of the rate of genome duplication
to the rate of gene loss.  The behavior we obtain for all species
indicates that the rate of genome duplication, relative to the
rate of gene loss, is approximately the same for each species.
This points to the ancient origin of the cellular machinery
responsible for the duplication of DNA.

Different classes of genes evolve at slightly different rates.
Families that perform cellular processes tend to be larger than
average.  Supplementing these functions might provide a
disproportionate selective advantage.  Also, the remaining
functions (information storage and processing and metabolism)
could represent core cellular machinery that is relatively
standard and requires less variability.

It would be interesting to verify whether the same protein family
size distributions are observed in multicellular plants and
animals.  One might expect that genome duplication would be
supplanted by chromosome duplication, which would shift the family
size distribution from a power law to a steeper, almost
exponential decay.  Some evidence in this direction is already
provided with the S. cerevisiae data presented in
Sec.~\ref{s:results}. With the C.~elegans sequence
reported~\cite{celegans}, the D.~melanogaster sequence promised
within a year~\cite{drosophila}, and a rough draft of the
H.~sapiens genome imminent~\cite{homosapiens}, this question might
soon be answered.

\acknowledgments

We wish to thank Piotr Slonimksi for drawing our attention to
models of genome evolution and Eugene Koonin for providing
technical assistance with the COG database.

\begin{figure}
\caption{ The family size distribution $P_n/P_1$ is shown for
three values of $\alpha$: 0.1 (thin line), 0.3 (medium line), and
0.9 (thick line).  Results are displayed both for Model I (grey
lines) and Model II (black lines).  Model II, which predicts a
power-law distribution, is linear on a log-log plot. }
\label{f:model}
\end{figure}

\begin{figure}
\caption{The size distributions $P_n/P_1$ of protein families are
shown for the eight organisms included in the COG database.  The
linear trend on the log-log plot is evidence for genome
duplication being the primary evolutionary mechanism driving the
growth of gene families.  Lines are provided as a guide to the
eye. } \label{f:all}
\end{figure}

\begin{figure}
\caption{The family size distributions $P_n/P_1$ are shown for
protein families in E. coli (left half) and S. cerevisiae (right
half).  Also shown are predictions of Model I (gene duplications
are independent, grey line) and Model II (the entire genome
duplicates, black line).} \label{f:fitorg}
\end{figure}

\begin{table}
\caption{The parameter $\alpha$ as calculated from Model I and
Model II is presented, along with the RMS of the fit, for the
organisms and functional categories in the COG database.
\label{t:alpha}}
\begin{tabular}{cddddrc} \\
\multicolumn{1}{l}{Organism / Functional category} &
\multicolumn{2}{c}{Model I} & \multicolumn{2}{c}{Model II} &
$N_{\rm fit}$\tablenote{$N_{\rm fit}$ is the number of family
sizes used in the fit (all sizes with 2 or more families).} &
Better Model
\\
& $\alpha$ & RMS  & $\alpha$ & RMS & &
\\
\\
 \multicolumn{1}{l}{E. coli / All} &  0.84    &   0.39 & 0.50 &
0.16 & 17 & II
\\
Information\tablenote{Information storage and processing}  &
0.77 & 0.55 & 0.47 & 0.39 & 6 & II
\\
Cellular processes  &   0.84 &   0.20 & 0.66 & 0.12 &   7 &   II
\\
Metabolism & 0.81    &   0.32    & 0.53 & 0.17 & 10  &   II
\\
Poorly characterized &   0.89    & 0.39 & 0.64 & 0.25 &   8 &   II
\\ \\
\multicolumn{1}{l}{H. influenzae /All} & 0.56 & 0.22 & 0.31 & 0.09
& 8 & II
\\
Information  & 0.36    & 0.10 & 0.25 & 0.12 & 4 & I
\\
Cellular processes &   0.73 & 0.14    & 0.54    & 0.03 & 4 &   II
\\
Metabolism   & 0.53 &   0.16 &   0.34    & 0.04    & 5 & II
\\
Poorly characterized &   0.56    & 0.10 &   0.41 &   0.05 & 5 & II
\\ \\
\multicolumn{1}{l}{H. pylori / All} & 0.54 & 0.32 & 0.30 & 0.13 &
7 & II
\\
Information & 0.47 &   0.25    & 0.30    & 0.14 & 5 & II
\\
Cellular processes &   0.48 & 0.01    &   0.38 &   0.09    & 4 & I
\\
Metabolism &   0.33    & 0.15    & 0.26 &   0.08 & 3   & II  \\
Poorly characterized &   0.73 & 0.39    & 0.49 & 0.25    & 5   &
II
\\ \\
\multicolumn{1}{l}{M. genitalium / All}  &   0.21 & 0.14 & 0.15 &
0.05 & 3 & II
\\
Information &   0.11 & 0.00    & 0.11    & 0.00 & 2 & Tie
\\
Cellular processes & 3.12 &   0.00    & 3.12 & 0.00    &   1   &
Tie
\\
Metabolism & 0.12    & 0.00    &   0.12 &   0.00    &   2   & Tie
\\
Poorly characterized    & 0.39    &   0.08 & 0.32    & 0.03 & 3 &
II
\\ \\
\multicolumn{1}{l}{M. jannaschii / All}  & 0.75 & 0.57 & 0.41 &
0.31 & 7 & II
\\
Information & 0.54    &   0.18    & 0.42 & 0.13 & 4 & II
\\
Cellular Processes &   0.70 & 0.03    & 0.62    &   0.07    &   4
& I
\\
Metabolism & 0.53 &   0.21 &   0.34    &   0.08    &   5 & II
\\
Poorly characterized & 0.64    &   0.07    &   0.56    & 0.11    &
4 & I
\\ \\
\multicolumn{1}{l}{M. pneumoniae / All}  &   0.26 & 0.18 & 0.19 &
0.10 & 3 & II
\\
Information  &   0.20    &   0.09    & 0.15 & 0.00 & 3 & II
\\
Cellular processes &   0.42 &   0.00    & 0.33    &   0.00 &   2 &
Tie
\\
Metabolism &   0.23    & 0.22    &   0.17    & 0.14    & 3 &   II
\\
Poorly characterized &   0.39    &   0.08    & 0.32    & 0.03 & 3
& II
\\ \\
\multicolumn{1}{l}{Synechocystis / All}   &   0.73 & 0.37 & 0.42 &
0.14 & 10 & II
\\
Information   &   0.49    & 0.18    & 0.33 & 0.09 & 5 & II
\\
Cellular processes  & 0.83 &   0.08    &   0.70 & 0.12 & 6   &   I
\\
Metabolism & 0.54    &   0.23    &   0.32 & 0.08 & 6 &   II
\\
Poorly characterized & 0.77    &   0.27 & 0.55    & 0.15 &   8 &
II
\\ \\
\multicolumn{1}{l}{S. cerevisiae / All} & 0.82 & 0.25 & 0.57 &
0.17 & 12 & II
\\
Information   &   0.84 &   0.20    & 0.76 & 0.22 & 6 & I
\\
Cellular processes  &   0.95    & 0.16 & 0.92    & 0.16 &   5 & I
\\
Metabolism &   0.72    & 0.14    & 0.50 &   0.10 & 9   &   II   \\
Poorly characterized & 0.95 &   0.13    & 0.85    & 0.09 &   8   &
II
\\ \\
\end{tabular}
\end{table}

\begin{table}
\caption{The number of organisms for which Model I or Model II is
a better fit is summarized according to protein functional
classes. \label{t:summary}}
\begin{tabular}{cccc}

Functional class & Model I Better & Model II Better & Tie \\
\tableline

All classes & 0 & 8 & 0
\\
Information\tablenote{Information storage and processing} & 2 & 5
& 1
\\
Cellular processes & 4 & 2 & 2
\\
Metabolism & 0 & 7 & 1
\\
Poorly characterized & 1 & 7 & 0
\\
\end{tabular}
\end{table}

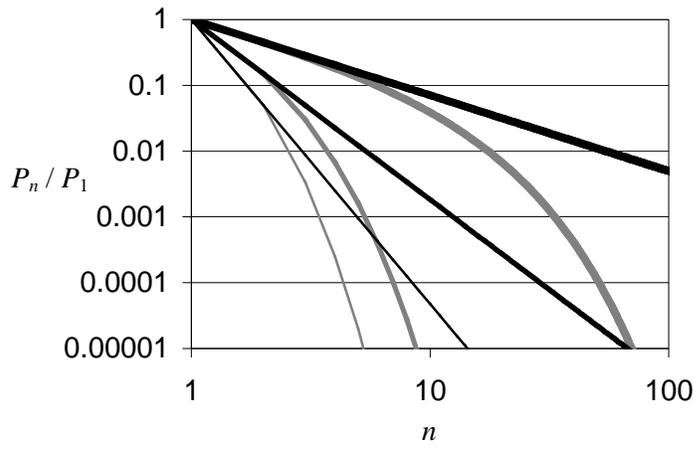
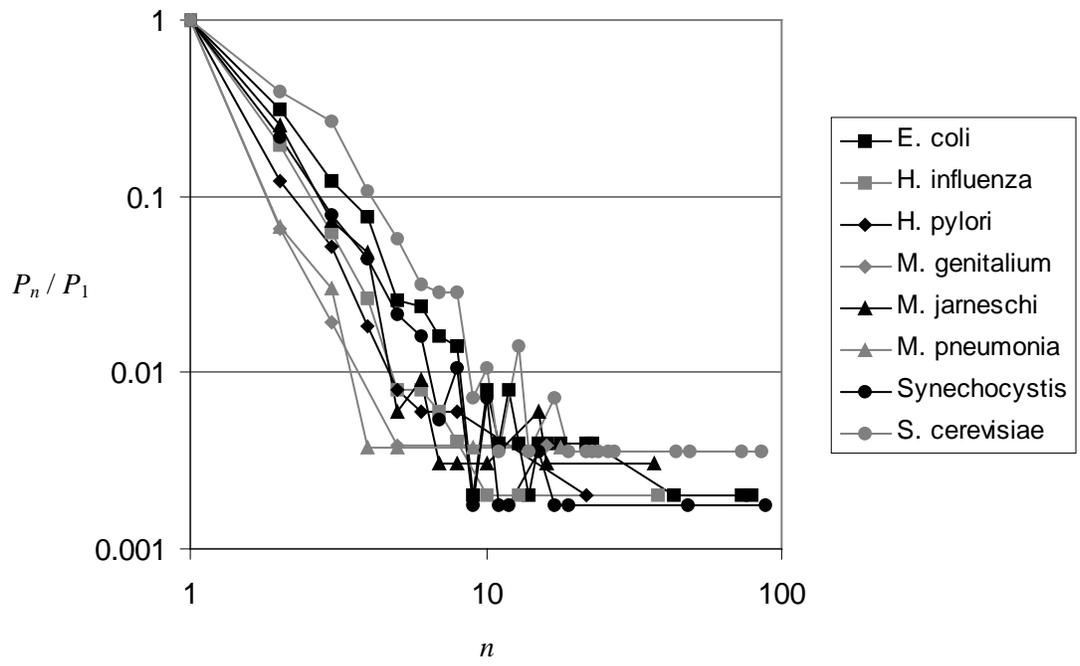
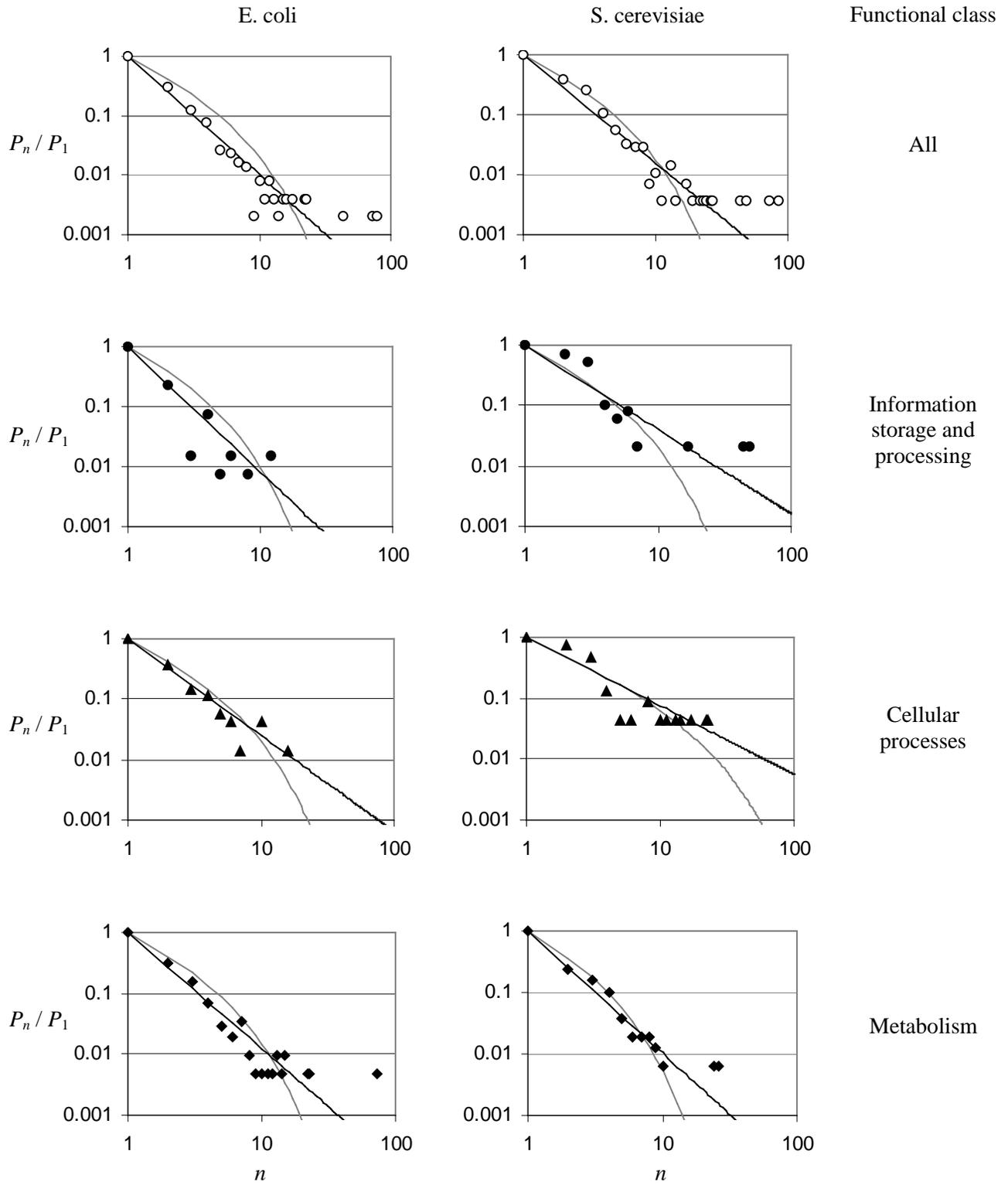
\end{document}